\documentclass[preprint,5p,times,twocolumn]{elsarticle}

\usepackage{epstopdf}
\usepackage{widetext}
\usepackage{amsmath}
\usepackage{varioref}
\usepackage{xr-hyper}
\usepackage{hyperref}
\usepackage{multirow}

\usepackage{color}
\usepackage{multirow}
\usepackage{hhline}
\usepackage{tabularx}
\usepackage[normalem]{ulem}

\hypersetup{colorlinks=true,citecolor=blue,linkcolor=blue,filecolor=blue,urlcolor=blue}

\def\be{\begin{equation}}
\def\ee{\end{equation}}
\def\ba{\begin{array}}
\def\ea{\end{array}}
\def\bea{\begin{eqnarray}}
\def\eea{\end{eqnarray}}

\usepackage{epsfig}

\usepackage{amssymb}

\journal{Physics Letters B}

\begin{document}

\begin{frontmatter}



\title{Structure effects on fission yields}


\author{Bharat Kumar$^{a,b}$}
\author{M.T. Senthil kannan$^{c}$}
\author{M. Balasubramaniam$^{c}$}
\author{B. K. Agrawal$^{b,d}$}
\author{S. K. Patra\corref{SKP}$^{a,b}$}
\cortext[SKP]{Corresponding author}
\ead{patra@iopb.res.in}
\address{$^{a}$Institute of Physics, Sachivalaya Marg, Bhubaneswar - 751005, India.}
\address{$^{b}$Homi Bhabha Nnational Institute, Anushakti Nagar, Mumbai - 400094, India.}
\address{$^{c}$Department of Physics, Bharathiar University, Coimbatore - 641046, India.}
\address{$^{d}$Saha Institute of Nuclear Physics, 1/AF,  Bidhannagar, Kolkata - 700064, India.}

\begin{abstract}

The structure effects of the fission fragments on their yields
are studied within the statical theory with the inputs, like, excitation
energies and level density parameters for the fission fragments at a given
temperature calculated using the temperature dependent relativistic
mean field formalism (TRMF).  For the comparison, the results are also
obtained using the finite range droplet model. At temperatures $T =1-2$
MeV, the structural effects of the fission fragments influence their
yields. It is also seen that at
$T = $ 3 MeV, the fragments become spherical and the fragments distribution
peaks at a close shell or near close shell nucleus.

\end{abstract}


\end{frontmatter}


\section{Introduction}
\label{sec1}
Nuclear fission is still not completely understood, although it was
discovered eight decades ago. The study of fission mass distribution
is one of the major insight of the fission process. Conventionally,
there are two different approaches, the statistical and the dynamical
approaches for the study of fission process \cite{gre73,fon56}. The
latter is a collective calculations of the potential energy surface
and the mass asymmetry. Further, the fission fragments were determined
either as the minima in the potential energy surface or by the maximum in
the WKB penetration probability integral for the fission fragments. The
statistical theory \cite{fon56}, begins with the statistical equilibrium
is established from the saddle to scission point and the relative
probability of the fission process depends on the density of the
quantum states of the fragments at scission point. The mass and the
charge distribution of the binary and the ternary fission is studied
using the single particle energies of the finite range droplet model
(FRDM) \cite{mrd81,mbs2014,sk16}. In FRDM \cite{moller95} formalism,
the energies at given temperature  are calculated using the relation $E(T) = \sum_i
n_i \epsilon_i$  with $n_i$ and $\epsilon_i$ are the fermi-distribution
function  and the single-particle energy corresponding to the ground
state deformation \cite{mbs2014}. 
The temperature dependence of the deformations of the fission fragments and  the
contributions of the pairing correlations are also ignored.
In the present letter, we applied
the self consistent temperature dependent relativistic mean field theory
(TRMF) with the well known NL3 parameter set \cite{bka00,mts16}. Here,
we study the structure effects of the fission fragments, whether they
could influence the yield or not ? If it influences, to what extent of
the temperature it affects the probable fragmentation.

The letter is organized as follows. In Section \ref{sec2}, we present a
 brief description of the statistical theory and the TRMF formalism. 
In Section \ref{sec3}, we discuss the variation of the quadrupole deformation
parameter $\beta_2$ of the fragments with the temperature and study the 
structure effects of the fission fragments in the mass distribution. The summary 
and conclusions of our results are given in Section \ref{sec4}.

\section{The method} \label{sec2}
We constraint the charge to mass ratio of the fission fragments
to that of the parent nucleus 
i.e.,  $Z_P/A_P \approx Z_i/A_i$, with $A_P$, $Z_P$ and $A_i$, $Z_i$ 
($i$ = 1 and 2) correspond to mass and
charge number of the parent nucleus and fission fragments, respectively
\cite{mrd81}. The constraints, $A_1 + A_2 = A$ and $Z_1 + Z_2 = Z$ are
imposed to account for  the mass and charge conservations. 
For $^{242}$Pu, which is a representative case in the present study, 
the binary charge numbers are restricted to $ Z_2 \ge $ 26 and $Z_1 \le$
66 by considering the experimental yield \cite{berg71}.
According to statistical theory \cite{fon56,cole}, the
probability of the particular fragmentation is directly proportional to
the folded level density $\rho_{12}$ of that fragments with the total
excitation energy $E^*$.
\begin{equation}
\rho_{12}(E^*) = \int_{0}^{E^*} \rho_{1}(E^*_{1})\,\rho_{2}(E^* - E_1^*)\, dE^*_1.
\end{equation} \label{eq1}

The folded level density is calculated from the fragment level densities 
$\rho_i$ \cite{mrd81,sk16} with the excitation energy $E_i^*$ and the level 
density parameters $a_i$ ($i$ = 1 and 2), which is given as:
\begin{equation}
\rho_i\left(E^{\ast}_i\right) = \dfrac{1}{12} \left(\dfrac{\pi^{2}}{a_i}\right)^{1/4} 
E^{\ast (-5/4)}_i\exp\left(2\sqrt{a_iE^{\ast}_i}\right).\label{eq2}
\end{equation}

Here, we 
used the self consistent temperature dependent 
relativistic mean field theory \cite{bka00,mts16} to calculate the 
excitation energy of the fragments. The excitation energies are 
calculated using the
total energy for a given temperature and the ground 
state energy ($T =$ 0)  as, $ E^{*}_{i} = E(T) - E(T = 0) $. Further, 
we calculate the excitation energies of the fragments using the 
ground state single particle energies of Finite Range Droplet Model 
(FRDM) 
\cite{moller95}  at the given temperature $T$ with 
the help of the Fermi-Dirac distribution function keeping
the total particle number conserved. The level density parameters 
$a_i$ are evaluated from the excitation energies and the temperature 
as $a_i = E^*_i/T^{2}$. The relative yield is estimated as the ratio 
between the probability of a given fragmentation and the sum 
of the probabilities of all the possible fragmentations and it is given by,
\begin{equation}
Y(A_j,Z_j)=\frac{P(A_j,Z_j)}{\sum P(A_j,Z_j)}. \label{eq3}
\end{equation}
Here, the sum of the total probability is normalized to the scale 2.
The competing basic decay modes such as neutron emission and $\alpha$ decay 
are not considered and the presented results are the prompt disintegration 
of a parent nucleus into two fragments (democratic breakup).

The well known relativistic mean field theory considers the nucleon 
nucleon interaction via effective mesons. The nucleon-meson interaction 
Lagrangian density are found in \cite{gam90,patra91,mts16}. The TRMF 
formalism is explained in Refs. \cite{bka00,mts16}. Here, we briefly 
revisit the RMF with the BCS formalism. For open shell nuclei the pairing 
is important for the shell structure. The constant pairing gap BCS approximation 
is used in our calculations. The pairing interaction energy in terms of 
occupation probabilities $v_i^2$ and $u_i^2=1-v_i^2$ is written 
as~\cite{pres82,patra93}:
\begin{equation}
E_{pair}=-G\left[\sum_{i>0}u_{i}v_{i}\right]^2,
\end{equation}
with $G$ is the pairing force constant. The variational approach with 
respect to the occupation number $v_i^2$ gives the BCS equation 
$2\epsilon_iu_iv_i-\triangle(u_i^2-v_i^2)=0$ with the pairing gap 
$\triangle=G\sum_{i>0}u_{i}v_{i}$. The pairing gap ($\triangle$) of proton 
and neutron is taken from the empirical formula\cite{gam90}, $\triangle = 
12 \times A^{-1/2}$ MeV. The temperature is introduced in the partial 
occupancies in the BCS approximation as: 
\begin{equation}
n_i=v_i^2=\frac{1}{2}\left[1-\frac{\epsilon_i-\lambda}{\tilde{\epsilon_i}}[1-2 f(\tilde{\epsilon_i},T)]\right],
\end{equation}
with the function $f(\tilde{\epsilon_i},T) = 1/(1+exp[{\tilde{\epsilon_i}/T}])$ 
represents the Fermi Dirac distribution; $\tilde{\epsilon_i} = \sqrt{(\epsilon_i-\lambda)^2+\triangle^2}$ 
is the quasi particle energies and the $\epsilon_i$ is the single particle energies. In lower 
temperatures, the pairing energies play an important role in the structure of open shell nuclei. The chemical potential $\lambda_p (\lambda_n)$ for protons (neutrons) is obtained from the particle number constraints $\sum_i n_i^{Z} (n_i^{N})  = Z(N) $. The sum is taken over all proton and neutron states. The entropy is obtained by,
\begin{equation}
S = - \sum_i \left[n_i\, ln n_i + (1 - n_i)\, ln (1- n_i)\right].
\end{equation}
The temperature dependent RMF binding energies and the gap parameter are obtained by minimizing the free energy,
\begin{equation}
F = E - TS.
\end{equation}

In constant pairing gap calculation, for a particular value of pairing gap $\triangle$ and force constant $G$, the pairing energy $E_{pair}$ diverges, if it is extended to an infinite configuration space. Therefore, a pairing window in all the equations are extended up-to the level $|\epsilon_i-\lambda|\leq 2(41A^{-1/3})$ as a function of the single particle energy \cite{gam90,patra93}.

\begin{figure}[!b]
	\includegraphics[width=1\columnwidth]{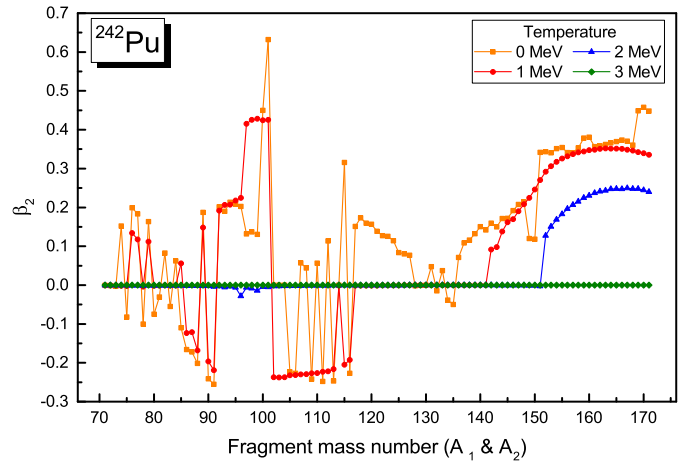}
	\caption{(color online) The variation of the quadrupole deformation parameter
 $\beta_2$ for the fission fragments for $^{242}$Pu at temperature $T =$ 0-3 MeV.}
	\label{deform}
\end{figure}
\begin{figure*}[!t]
	\includegraphics[width=2\columnwidth]{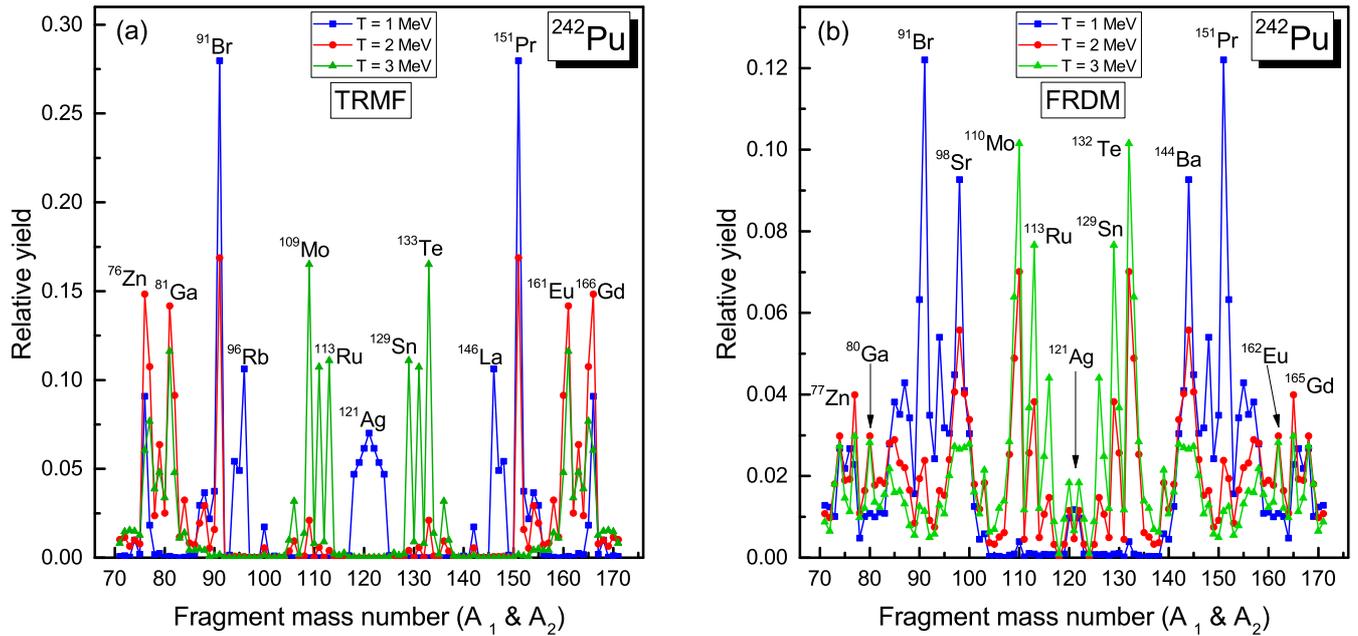}
	\caption{(color online) Mass distribution of $^{242}$Pu for the three different temperatures $T =$ 1, 2 and 3 MeV 
using the TRMF and FRDM formalisms. The sum of the total yield is
normalized to the scale 2.}
	\label{massdis}
\end{figure*}

\section{Results and discussions} \label{sec3}

In this letter we study the fission fragment distributions of $^{242}$Pu
as a representative case. After generating the possible fragments by
equating the charge to mass ratio of the parents to the charge-mass ratio
of the fragments, the yield values are obtained using Eq. \ref{eq3}. The
folded densities are calculated from the individual level density of
the fragments which are evaluated from the fragment excitation energies
at the given temperature $T$.  The excitation energies are calculated
within the TRMF and FRDM. In TRMF model, the excitation energy is
the difference between the temperature dependent total energy 
and the ground state 
energy.  The minimization
of free  energy of fission fragments within TRMF are obtained by allowing
them to be axially deformed.  The solutions for the Dirac equations for
nulceons and the Klein Gordon equations for the meson fields are obtained
by expanding them in terms of the spherical harmonic wave functions as
basis. For the numerical calculations, the nucleon wave functions and
the meson fields are expanded to the number of major shells taken to be
$N_F=12$   and $N_B=20$, respectively.
In the FRDM method, the excitation energies
are calculated using the single particle energies which are retrieved
from the Reference Input Parameter Library (RIPL-3) \cite{ripl3} and
the details of calculations can be found in Refs. \cite{mbs2014,sk16}.

\begin{table}
	\caption{The relative fission yield (R.Y.)=$Y(A_j,Z_j)=\frac{P(A_j,Z_j)}{\sum P(A_j,Z_j)}$ obtained at temperature $T= $1, 2 and 3 MeV from the TRMF calculations are 
		compared with the FRDM prediction for $^{242}$Pu. The sum of the total yield is
normalized to the scale 2.} \label{table} 
	\centering
	\renewcommand{\arraystretch}{1.2}
	\begin{tabular}{ccccc}
		
		\hline
		\hline
		T (MeV) & \multicolumn{2}{c}{TRMF } & \multicolumn{2}{c}{FRDM }\\
		\hline
		
		&Fragment &R.Y.& Fragment & R.Y.\\
		\hline
		& $^{91}$Br + $^{151}$Pr &0.559& $^{91}$Br + $^{151}$Pr & 0.244\\
		
		&  $^{96}$Rb + $^{146}$La &0.213&  $^{98}$Sr + $^{144}$Ba &0.184\\
		
		&  $^{76}$Zn + $^{166}$Gd & 0.182& 	$^{90}$Br + $^{152}$Pr&0.120 \\
		
		1 &  $^{121}$Ag + $^{121}$Ag & 0.140& $^{94}$Rb + $^{148}$La& 0.108 \\  
		
		& $^{120}$Ag + $^{122}$Ag & 0.122& $^{97}$Sr + $^{145}$Ba &0.088 \\
		
		&  $^{94}$Rb + $^{148}$La & 0.108& $^{87}$Se + $^{155}$Nd&0.086 \\ 	
		
		&  $^{119}$Pd + $^{123}$Cd & 0.107& $^{99}$Sr + $^{143}$Ba &0.082  \\	
		
		&  $^{95}$Rb + $^{147}$La & 0.098& $^{85}$As + $^{157}$Pm&0.076 \\	
		
		\hline
		& $^{91}$Br + $^{151}$Pr &0.336& $^{110}$Mo + $^{132}$Te &0.140\\
		&  $^{76}$Zn + $^{166}$Gd & 0.295& $^{98}$Sr + $^{144}$Ba &0.110\\
		
		&  $^{81}$Ga + $^{161}$Eu &0.282& $^{97}$Sr + $^{145}$Ba&0.080 \\
		
		2 	&$^{77}$Zn + $^{165}$Gd &0.213&  $^{99}$Sr + $^{143}$Ba &0.079\\
		
		&$^{82}$Ge + $^{160}$Sm &0.182&  $^{113}$Ru + $^{129}$Sn&0.076  \\
		
		&$^{79}$Ga + $^{163}$Eu &0.125& $^{100}$Y + $^{142}$Cs&0.068 \\
		
		\hline
		& $^{109}$Mo + $^{133}$Te &0.329& $^{110}$Mo + $^{132}$Te &0.202\\
		
		&   $^{81}$Ga + $^{161}$Eu &0.295&  $^{113}$Ru + $^{129}$Sn&0.152 \\
		
		&$^{113}$Ru + $^{129}$Sn &0.222&  $^{109}$Mo + $^{133}$Te&0.128 \\
		
		3	&   $^{77}$Zn + $^{165}$Gd &0.120& $^{116}$Rh + $^{126}$In&0.087 \\
		
		&  	$^{79}$Ga + $^{163}$Eu & 0.095& $^{112}$Ru + $^{130}$Sn&0.072 \\	
		
		&  $^{82}$Ge + $^{160}$Sm & 0.094&   $^{77}$Zn + $^{165}$Gd&0.059 \\
		
		\hline \hline
		\vspace{0.2cm}
	\end{tabular}
	
\end{table}

In  Fig. \ref{deform}  we plot the quadrupole deformation parameter
$\beta_2$ as a function of mass numbers $A_1$ and $A_2$ for the binary
fragments. 
For $T =$ 1 MeV, the quadrupole deformation parameter of the fragments are prolate/oblate or spherical similar to the ground state nuclei.  The pairing transition of
the nuclei occur within the temperature $T=1$ MeV. For $T =$ 2 MeV, the
fragments with mass number A $\le$ 150 become almost spherical except
few exceptions.  
For higher temperature $T =$ 3 MeV, all the nuclei become perfectly
spherical.  When the temperature increases, the levels above the Fermi
surface become more occupied due to the transition of particles from
the partial occupied levels (below the Fermi surface).  For higher
temperatures, the occupancies become uniform at the 
vicinity  of the Fermi surface. The deformed  structures are melted
at higher temperatures, and 
the evolved fragments land up in a close shell shape
due to the higher excitation energy of the closed shell nuclei.

Now the fission mass distributions of $^{242}$Pu obtained from the TRMF
and FRDM formalisms for three different temperatures $T =$ 1, 2 and 3 MeV
are shown in Fig. \ref{massdis} and the fragment yields are compared in
table \ref{table}. For $T =$ 1 MeV, the most favorable fragmentation is
$^{91}$Br $+ ^{151}$Pr, for  both the models. However, other fragments
such as $^{96}$Rb $+ ^{146}$La in TRMF and $^{98}$Sr $+ ^{144}$Ba in FRDM
are also noticed. In addition, the binary fragmentations $^{121,122}$Ag
$+ ^{121,120}$Ag are favorable for both models. At temperature $T =$ 2
and 3 MeV the yields of the minor fragments evolution in FRDM model have
increased. But, in TRMF model, the most favorable fragments are mainly
from the two regions, I (with mass $A\sim 71-91$ and $A\sim 171-151$)
and II ($A\sim 105-113$ and $137-129$). For $T =$ 2 MeV, region II is
less probable than  region I. The most favorable fragments are same as
those for  temperature $T =$ 1 MeV, along with one of the neutron closed
shell nucleus in $^{81}$Ga + $^{161}$Eu (for $^{81}$Ga, N=50). Analyzing
Fig. 2(a) for TRMF results, we notice a symmetric yield region for
$T =$ 3 MeV, i.e. from mass region $A\sim 100-133$. Most of the yield
fragments are concentrated in this region. Additionally, 
a few more secondary asymmetric fractions are also
noticed. We see a similar symmetric region along with few asymmetric
yields in case of FRDM calculations also (see Fig. 2(b)). The only
difference is the secondary asymmetric yields spread widely in this case,
contrary to the concentrated symmetric peaks for TRMF calculations. In
general, our results agree with the recent experimental observations of
Ref. \cite{cha2015}.

In FRDM model, the favorable fragments are in the vicinity of the
neutron closed shell (N $=$ 82) nuclei or exactly at the proton
closed shell nuclei (Z $=$ 50). The other secondary fragmentations are
not necessarily closed shell nuclei, so a good fraction of non-magic
fragments are also appeared in the yield values.  For higher temperature
$T =$ 3 MeV, the most favorable fragments for both models are more or
less same. The fragment combinations $^{109,110}$Mo + $^{133,132}$Te
and $^{113}$Ru + $^{129}$Sn are the most favorable fragments. In TRMF
model the most favorable fragments in  region I and region II have
sharp peak yield values. But in FRDM, the larger yields are at region II
only. Conclusively, for $T =$ 3 MeV, one of the most favorable fragments
are at the vicinity of the closed shell (N $\sim 82$) or exactly at the
closed shell (N or Z = 50) nuclei.  As it is reported in many earlier
works, that the rare earth nuclei in the range $Z = 56-66$ region shows,
peculiar nature, such as shell closure behavior, most of the fragments
are found in that region. Also, the corresponding neutron
numbers for such nuclei act like deformed shell closure at $N = 98 -
102$ region. As a result, we get many neutron fragment in that region
\cite{sat04}.  Although, the even-even  fragments are more possible fission
yield, in the present case, we find maximum number of fission yield is
odd mass fragments as compared to the even-even combination. Because
of the level density of the odd mass fragments are higher than the even
mass fragments as reported in Ref. \cite{fon56}. For this temperature,
all nuclei become spherical, the excitation energy of the closed shell
nuclei is higher than the other spherical nuclei. Thus, the deformation
of the fission fragments affects the most favorable distribution at
the temperatures $T$ = 1-2 MeV. In FRDM model the temperature dependent
deformations are not considered, so the most favorable fragments at $T$
= 2 and 3 MeV are same.

\section{Summary and Conclusions}\label{sec4}

The binary fission mass distribution is studied within the statistical
theory.  The level densities and the excitation energies
are calculated from the TRMF and FRDM formalisms.  The TRMF model
includes the thermal evolutions of the pairing gaps and deformation
in a self-consistent manner, while, they are ignored in the FRDM.
The excitation energies of the fragments are obtained from ground state
results in the FRDM calculations.  So, the study of deformation effects
on the fission fragments can be seen within the TRMF formalism. The
quadrupole deformations of fission fragments with the increasing
temperature are also discussed. The structural effects of the fission
fragments influence the most favorable  yields at temperatures
$T$ = 1 and 2 MeV. For the temperature $T =$ 3 MeV,  one of the most
favorable mass distribution ends up in the vicinity or exactly at the
nucleon closed shell (N$\sim$82) or (N or Z = 50). Although, the TRMF and
FRDM formalisms  yield the closed shell nucleus as one of the favorable
fragment,  the detail of the fission yields and probability of the mass
distributions are quite different in these two formalisms.

The author MTS acknowledge that the financial support from UGC-BSR
research grant award letter no. F.25-1/2014-15(BSR)7-307/2010/(BSR)
dated \,\, 05/11/2015 and IOP, Bhubaneswar for the warm hospitality
and for providing the necessary computer facilities.





\end{document}